# Quantal self-consistent cranking model for monopole excitations in even-even light nuclei

P Gulshani

NUTECH Services, 3313 Fenwick Crescent, Mississauga, Ontario, Canada L5L 5N1
Tel. #: 905-569-8233; matlap@bell.net

**Abstract**

In this article, we derive a quantal self-consistent time-reversal invariant parameter-free cranking model for isoscalar monopole excitation coupled to intrinsic motion in even-even light nuclei. The model uses a wavefunction that is a product of monopole and intrinsic wavefunctions and a constrained variational method to derive, from a many-particle Schrodinger equation, a pair of coupled self-consistent cranking-type Schrodinger equations for the monopole and intrinsic systems. The monopole co-ordinate used is the trace of the quadrupole tensor and hence describes the overall deformation of the nucleus. The monopole and intrinsic wavefunctions are coupled to each other by the two cranking equations and their associated parameters and by two constraints imposed on the intrinsic system. For an isotropic Nilsson shell model and an effective residual two-body interaction, the two coupled cranking equations are solved in the Tamm Dancoff approximation. The strength of the interaction is determined from a Hartree-Fock self-consistency argument. The excitation energy of the first excited $0^+$ state is determined and found to agree closely with those observed in the nuclei ${}^{4}_{2}He$, ${}^{8}_{4}Be$, ${}^{12}_{6}C$, ${}^{16}_{8}O$, ${}^{20}_{10}Ne$, ${}^{24}_{12}Mg$, and ${}^{28}_{14}Si$. The variation of the model parameters are explained. In particular, it is found that the monopole excitation energy as a function of the mass number undergoes an increase whenever the nucleons begin to occupy a new sub-shell state with non-zero orbital angular momentum as a consequence of suppressing or constraining the resulting spurious monopole excitation in the intrinsic system.

## 1. Introduction

Collective excitations in nuclei, characterized by excited states with specific excitation energy, angular momentum, parity, and enhanced electromagnetic transition rates, have been studied theoretically using various methodologies ([1-13] and references in [7,9]). In particular, reference [1] used the phenomenological Inglis' cranking model with the dilation or compression cranking operator to study monopole oscillations in nuclei. References [2-6] used the nuclear radius as the monopole co-ordinate and expressed nuclear wavefunction as a series in either $K$ or hyperspherical harmonics to describe the properties of the nuclear monopole states and giant monopole resonances. Reference [7] used the nuclear radius in two spatial dimensions and a product wavefunction to transform a many-particle Schrodinger equation and obtain coupled monopole-intrinsic Schrodinger equations, which were approximately solved using their underlying $su(1,1)$ dynamical algebra. References [8,9] generalized the model in [7] to three dimensions and used a constrained variational method to derive coupled monopole and intrinsic self-consistent cranking-type Schrodinger equations.

In this article, we generalize the model in [8,9] to include a two-body interaction and solve the resulting constrained monopole-intrinsic self-consistent cranking Schrodinger equations in the Tamm Dancoff approximation, and determine the excitation energy of the first excited $0^+$ state in the ${}^{4}_{2}He$, ${}^{8}_{4}Be$, ${}^{12}_{6}C$, ${}^{16}_{8}O$, ${}^{20}_{10}Ne$, ${}^{24}_{12}Mg$, and ${}^{28}_{14}Si$ nuclei. Some of these nuclei are deformed but the model is still used to describe oscillations in the overall deformation in these nuclei because the monopole coordinate used is the trace of the particle quadrupole tensor distribution, which describes the overall nuclear quadrupole deformation. In Section 2, we derive the model equations and the equations for the model parameters using the methodology presented in [8,9]. In Section 3, we determine the solutions of the monopole and intrinsic cranking Schrodinger equations including a residual two-body interaction using Tamm Dancoff approximation for the intrinsic state. Expressions for the model parameters are then derived. An equation for the monopole frequency is derived using an energy-weighted sum rule and the cranking equations. Equations for the interaction strength are derived using a Hartree-Fock self-consistency argument. Other parameters in the model are determined self-consistently by the model equations, and hence there are no adjustable parameters in the model. The excitation energy of the first excited $0^+$ state is then defined. In Section 4, we use the model to compute the excitation energy of the first excited $0^+$ state in ${}^{4}_{2}He$, ${}^{8}_{4}Be$, ${}^{12}_{6}C$, ${}^{16}_{8}O$, ${}^{20}_{10}Ne$, ${}^{24}_{12}Mg$, and ${}^{28}_{14}Si$ nuclei, and compare them to the observed excitation energies. Section 5 presents concluding remarks.

## 2. Derivation of coupled self-consistent cranking monopole-intrinsic Schrodinger equations

We use the constrained variational method presented in [8,9] to derive coupled monopole-intrinsic self-consistent cranking Schrodinger equations as follows.

We start from the $A$-nucleon Schrodinger equation:

$$\left(-\frac{\hbar^2}{2M}\sum_{n,j=1}^{A,3}\frac{\partial^2}{\partial x_{nj}^2}+\hat{V}\right)\Phi=E\Phi \tag{1}$$

where the two-body interaction $\hat{V}$ and the wavefunction $\Phi$ are rotationally invariant (i.e., $\Phi$ is a state of zero total angular-momentum), and $x_{nj}$ $(n=1,\dots,A;\ j=1,2,3,$ where $A$ is the nuclear mass number) are space-fixed nucleon co-ordinates. We assume a product wavefunction of the form[1]:

$$\Phi=F(R)\cdot\phi(x_{ni}) \tag{2}$$

where the nuclear radius $R$, defined by:

[1] As is justified subsequently, the functions $F$ and $\phi$ are strongly coupled to each other by the two coupled cranking equations and their characterizing parameters, which are determined self-consistently by two cranking equations and two constraints on $\phi$.

$$R \equiv \sum_{n=1}^{A} r_n^2 \equiv \sum_{n=1}^{A} (x_n^2 + y_n^2 + z_n^2), \tag{3}$$

is assumed to be the monopole collective co-ordinate, and the intrinsic wavefunction $\phi$ (which can be identified with a shell-model wavefunction) is a function of $x_{nj}$. It is noted that *R* is the trace of the particle quadrupole tensor distribution and hence is proportional to the $\beta$ co-ordinate in the Bohr-Mottelson rotation-vibration model [17-20]. Therefore, oscillations in *R* correspond to oscillations in overall nuclear quadrupole deformation. $\phi$ is required to be independent of *R*, i.e., it is subject to the constraint:

$$\frac{\partial}{\partial R}\phi = 0, \tag{4}$$

and to be a zero total-angular-momentum state. Therefore, we require $\phi$ to be spherically symmetric even when we apply the model to deformed nuclei because the non-shape related overall nuclear deformation has been extracted and represented by the nuclear (monopole) radius *R*. Substituting Eq. (2) into Eq. (1) and using

Eqs. (3) and (4), we obtain:

$$-4R\frac{d^2F}{dR^2}\cdot\phi - 4\frac{dF}{dR}\cdot\tilde{B}\phi + F\cdot\left(-\sum_{n=1}^{A}\nabla_n^2 + \frac{2M}{\hbar^2}\hat{V}\right)\cdot\phi = \frac{2ME}{\hbar^2}F\,\phi \equiv \varepsilon F\,\phi \tag{5}$$

where $\tilde{B} \equiv \frac{1}{2}\sum_{n,j=1}^{A,3}\left(x_{nj}\frac{\partial}{\partial x_{nj}} + \frac{\partial}{\partial x_{nj}}x_{nj}\right)$ is the compression-dilation operator and $\varepsilon$ is the reduced energy defined on the right-hand side of Eq. (5).

We now write the two-body interaction $\hat{V}$ as the sum of its Hartree-Fock mean-field part $\hat{V}_{os}$, which is chosen to be the usual shell-model harmonic oscillator potential, and its remaining (residual) two-body part $\hat{V}_{mm}$, which is chosen to be the usual effective separable attractive monopole-monopole interaction:

$$\frac{2M}{\hbar^2}\hat{V} = \hat{V}_{os} + \hat{V}_{mm} == b^2\sum_{m=1}^{A} r_m^2 - \chi_m \sum_{n=1}^{A} r_n^2 \cdot \sum_{m=1}^{A} r_m^2 \tag{6}$$

where

$$b \equiv \frac{M\omega}{\hbar}, \qquad \hbar\omega = 41A^{-1/3}, \qquad \chi_m = \chi_{mo}/\Sigma, \qquad \Sigma \equiv \sum_{N=0}^{N_f}\left(N + 3/2\right) \tag{7}$$

(where the second expression in Eq. (7) is determined empirically [14]). In Eq. (7), $\Sigma$ is the total oscillator particle occupation number, *N* is an oscillator-shell quantum number, $N_f$ is the oscillator quantum number of the last particle-occupied (Fermi) shell, and the expression for the strength $\chi_m$ of $\hat{V}_{mm}$ is derived from a Hartree-Fock self-consistency and/or volume-conserving arguments [7,15,16]. We choose $\chi_{mo} = 0.124$, a value close to that commonly chosen for the constant $\chi_{mo}$.

We now define a restoring potential for the monopole oscillation by splitting each of $\hat{V}_{os}$ and $\hat{V}_{mm}$ into its monopole and intrinsic parts:

$$\hat{V}_{os} \equiv b^2 \cdot \sum_{m=1}^{A} r_m^2 \equiv b_m^2 \cdot R + b_s^2 \cdot \sum_{m=1}^{A} r_m^2 \tag{8}$$

$$\hat{V}_{mm} = -\chi_{mm} \cdot R^2 - \chi_{ms} \cdot \sum_{n=1}^{A} r_n^2 \cdot \sum_{m=1}^{A} r_m^2 \equiv -\chi_{mm} \cdot R^2 - \chi_{ms} \cdot V_{int}^2 \tag{9}$$

where the reduced monopole $b_m$, $\chi_{mm}$ and intrinsic $b_s$, $\chi_{ms}$ frequencies and interaction strengths are related by:

$$b^2 = b_m^2 + b_s^2, \qquad \chi_m = \chi_{mm} + \chi_{ms} \tag{10}$$

An expression for $\chi_{ms}$ (and hence for $\chi_{mm}$ through Eq. (10)) is given below in Eq. (16).

Following the procedure presented in [8,9], we substitute Eqs. (8) and (9) into Eq. (5), take the expectation of the resulting equation with request to the two wavefunctions in Eq. (2), subtract from the resulting equation the expectation of the constraint in Eq. (4)[2], use $\partial/\partial R = (6A)^{-1}\sum_{n,j=1}^{A,3}(x_{nj})^{-1}\partial/\partial x_{nj}$, and apply to the resulting equation the Rayleigh-Ritz variational method to obtain:

$$\left(-4R\frac{d^2}{dR^2} - 4a\frac{d}{dR} + b_m^2 R - \chi_{mm}\cdot R^2\right)|F\rangle = \left(\varepsilon - t_s - b_s^2\langle\phi|\sum_{n=1}^{A} r_n^2|\phi\rangle + \chi_{ms}\langle\phi|V_{int}^2|\phi\rangle\right)|F\rangle \tag{11}$$

$$\left(-\sum_{n=1}^{A}\nabla_n^2 + \beta_s\cdot\tilde{B} + b_s^2\sum_{n=1}^{A} r_n^2 - \chi\sum_{n,j=1}^{A,3}(x_{nj})^{-1}\partial/\partial x_{nj} - \chi_{ms}\cdot V_{int}^2\right)|\phi\rangle = \left(\varepsilon - t_R - b_m^2\langle F|R|F\rangle + \chi_{mm}\langle F|R^2|F\rangle\right)|\phi\rangle \tag{12}$$

$$a\cdot\beta_s = \varepsilon - t_s - t_R - b_m^2\langle F|R|F\rangle - b_s^2\langle\phi|\sum_{n=1}^{A} r_n^2|\phi\rangle + \chi_{mm}\langle F|R^2|F\rangle + \chi_{ms}\langle\phi|V_{int}^2|\phi\rangle \tag{13}$$

where $\chi$ is a Lagrange multiplier whose value is determined to satisfy the intrinsic constraint:

$$\langle\phi|\frac{\partial}{\partial R}|\phi\rangle = 0, \tag{14}$$

and where we have the definitions:

$$a \equiv \langle\phi|\tilde{B}|\phi\rangle,\ \beta_s \equiv -4\langle F|\frac{d}{dR}|F\rangle,\ t_s \equiv \langle\phi|\left(-\sum_{n=1}^{A}\nabla_n^2\right)|\phi\rangle,\ t_R \equiv \langle F|\left(-4R\frac{d^2}{dR^2}\right)|F\rangle,\ \tilde{B} \equiv \frac{1}{2}\sum_{n,j=1}^{A,3}\left(x_{nj}\frac{\partial}{\partial x_{nj}} + \frac{\partial}{\partial x_{nj}}x_{nj}\right) \tag{15}$$

We derive, similarly to that for $\chi_m$ in Eq. (7), the following expression for the intrinsic interaction strength $\chi_{ms}$:

$$\chi_{ms} = \frac{\chi_{mso} b_c b_s^2}{\bar{\Sigma}} \equiv \frac{\bar{\chi}_{mso}}{\Sigma}, \qquad \bar{\chi}_{mso} \equiv \frac{\chi_{mso} b_c b_s^2 \Sigma}{\bar{\Sigma}}, \qquad \bar{\Sigma} \equiv \Sigma + \sum_{l=0}^{l_f}\left[\sqrt{l(l+1) + \left(\frac{1+\chi}{2}\right)^2} - \left(l + \frac{1}{2}\right)\right] \tag{16}$$

using the method given in [7, Appendix $A$] and an $A$-particle system with the constrained Schrodinger equation:

$$H_c|\phi_o\rangle \equiv \left(-\sum_{n=1}^{A}\nabla_n^2 + \beta_s\cdot\tilde{B} + b_s^2\sum_{n=1}^{A} r_n^2 - \chi\sum_{n,j=1}^{A,3}(x_{nj})^{-1}\partial/\partial x_{nj}\right)|\phi_o\rangle = (\varepsilon_o)|\phi_o\rangle$$

which is Eq. (12) without the two-body interaction term $-\chi_{ms}\cdot V_{int}^2$. In Eq. (16), $l$ is the orbital angular momentum of the sub-shell single-particle state given by $N = 2n + l$, $N, n = 0,1,2,....\infty$, and $l = N, N-2,.....,1 \text{ or } 0$, and $l_f$ is the orbital angular momentum for the last occupied single-particle state. The expression in Eq. (16) is easily derived algebraically using the $su(1,1)$ dynamic algebra for $H_c$. In this article we choose $\chi_{mso}$ equal to 0.0423.

Simultaneous solution of the three Eqs. (11), (12), and (13), together with Eqs. (7), (10), (14), (15), and (16), determines the four parameters $\varepsilon$, $a$, $\beta_s$, and $\chi$ in terms of the monopole frequency $b_m$. Eq. (24) (given in Section 3) determine $b_m$ (and hence $b_s$ through Eq. (10)).

---

[2] That is, we are approximating the constraint in Eq. (4) by its first moment $\langle\phi|\frac{\partial}{\partial R}|\phi\rangle = 0$. It turns out that limiting the calculation to only this moment and ignoring the higher moments is sufficiently accurate.

Eqs. (11) and (12) may be viewed as microscopic, self-consistent, cranking Schrödinger equations for the monopole oscillations and the intrinsic motion respectively, with the cranking parameters $a$ and $\beta_s$ being dynamical variables determined self-consistently by Eqs. (11), (12), and (13). A consequence of this self-consistency is that Eqs. (11) and (12) are time-reversal invariant unlike that in the conventional phenomenological cranking models.

## 3. Solutions of Eqs. (11) and (12), and expressions for model parameters and excitation energy

In this article, we ignore for simplicity the anharmonic term $-\chi_{mm}\cdot R^2$ on the left-hand-side of Eq. (11). The eigenfunctions and eigenvalues of Eq. (11) are then obtained in closed form from the literature (as in [9]). We then evaluate, as in [9], the following expressions for the monopole-system quantities:

$$\varepsilon - t_s - b_s^2 \left\langle \phi \right| \sum_{n'=1}^{A} r_n^2 \left| \phi \right\rangle + \chi_{ms} \left\langle \phi \right| V_{int}^2 \left| \phi \right\rangle = 2 b_m \cdot (2n + a),\ t_R = b_m \cdot (2n + a - \mathcal{B}_n),\ \left\langle F \right| R \left| F \right\rangle = \frac{2n + a}{b_m},\ \left\langle F \right| R^2 \left| F \right\rangle = \frac{\bar{\mathcal{B}}_n}{b_m^2} \tag{17}$$

where the monopole quantum number $n = 0, 1, 2, \ldots.\infty$, and $\mathcal{B}_n = 2,\ \bar{\mathcal{B}}_n = a(a+1)$ for $n = 0$ and $\mathcal{B}_n = 2a,\ \bar{\mathcal{B}}_n = (a+1)(a+3)$ for $n = 1$.

The eigenfunctions and energy eigenvalues of Eq. (12) are obtained in two steps. First the independent-fermion Schrodinger equation:

$$H_o \left| \phi_o \right\rangle \equiv \left( -\sum_{n=1}^{A} \nabla_n^2 + \beta_s \cdot \tilde{B} + b_s^2 \sum_{n=1}^{A} r_n^2 \right) \left| \phi_o \right\rangle = \varepsilon^o \left| \phi_o \right\rangle \tag{18}$$

is solved to determine the ground-state Slater-determinant $\left| \phi_o \right\rangle$, as an anti-symmetrized product of harmonic-oscillator-related eigenfunctions (i.e., particle orbitals), the single-particle energy eigenvalues $\varepsilon_{Nnl}^o$, and the total ground-state energy $\varepsilon^o$:

$$\varepsilon_{Nnl}^o = 2\sqrt{b_s^2 + \beta_s^2} \cdot \left( N + \frac{3}{2} \right),\quad \varepsilon^o = 2\sqrt{b_s^2 + \beta_s^2} \cdot \varSigma \tag{19}$$

where $\varSigma$ is defined in Eq. (7). Then, we solve the constrained and interacting-particle Schrodinger equation:

$$\begin{aligned} \left( H_o - \chi \sum_{n,j=1}^{A,3} \frac{1}{x_{nj}} \frac{\partial}{\partial x_{nj}} - \chi_{ms} \cdot V_{int}^2 \right) \left| \phi \right\rangle &\equiv \left( H_o - \chi D - \chi_{ms} \cdot V_{int}^2 \right) \left| \phi \right\rangle \\ &= \left( \varepsilon - t_R - b_m^2 \left\langle F \right| R \left| F \right\rangle + \chi_{mm} \left\langle F \right| R^2 \left| F \right\rangle \right) \left| \phi \right\rangle \equiv \bar{\varepsilon} \left| \phi \right\rangle \end{aligned} \tag{20}$$

using Tamm Dancoff approximation [17-20] to expand $\left| \phi \right\rangle$ in a series of particle-hole excited states $\left| n j^{-1} \right\rangle$ obtained from a single-particle excitation of the "Hatree-Fock" vacuum $\left| \phi_o \right\rangle$:

$$\left| \phi \right\rangle = \sum_{nj} C_{nj} a_n^\dagger a_j \left| \phi_o \right\rangle \equiv \sum_{nj} C_{nj} \left| n j^{-1} \right\rangle \tag{21}$$

where $a_n^\dagger$ and $a_j$ are respectively the fermion creation and annihilation operators and the integers $n$, $m$ indicate particle states (above the Fermi surface) and $i, j$ indicate hole states (below the Fermi surface). $\left| \phi \right\rangle$ is subject to the normalization condition $\sum_{nj} C_{nj}^2 = 1$ and the constraint in Eq. (14), and must be a zero angular momentum state[3]. Substituting Eq. (21) into Eq. (20), and

[3] This condition restrict admissible types of particle-hole excitations.

multiplying the resulting equation by $\langle mi^{-1}|$, we obtain the matrix equation:

$$\begin{aligned}&\left(\varepsilon^{o}_{mi}-\chi\cdot\langle mi^{-1}|D|mi^{-1}\rangle-\chi_{ms}\cdot\langle mi^{-1}|V^{2}_{\text{int}}|mi^{-1}\rangle-\Delta\varepsilon\right)\cdot C_{mi}\\&\qquad-\sum_{nj\neq mi}\left(\chi\cdot\langle mi^{-1}|D|nj^{-1}\rangle+\chi_{ms}\cdot\langle mi^{-1}|V^{2}_{\text{int}}|nj^{-1}\rangle\right)\cdot C_{nj}=0\end{aligned}\tag{22}$$

where $\varepsilon^{o}_{mi}\equiv\varepsilon^{o}_{m}-\varepsilon^{o}_{i}$, $\varepsilon^{o}_{m},\varepsilon^{o}_{i}\equiv\varepsilon^{o}_{N\,nl}$ (refer to the definitions in Eq. (19)) and $\Delta\varepsilon\equiv\bar{\varepsilon}-\varepsilon^{o}$ ( refer to the definitions in Eqs. (19) and (20)).

For an appropriate choice of the particle-hole configuration space of a finite dimension, Eq. (22), together with Eqs. (14) and (16) and the normalization condition on $|\phi\rangle$, is solved for $\chi$, $C_{mi}$, $\chi_{ms}$, and $\Delta\varepsilon$. We then evaluate, as in [9], the following intrinsic-system quantities appearing in Eqs. (11), (12), and (13):

$$\bar{\varepsilon}\equiv\varepsilon-t_{R}-b_{m}^{2}\langle F|R|F\rangle+\chi_{mm}\langle F|R^{2}|F\rangle=\varepsilon^{o}+\Delta\varepsilon,\ \ \langle\phi|\sum_{n=1}^{A}r_{n}^{2}|\phi\rangle,\ \ t_{s}\equiv\langle\phi|\left(-\sum_{n=1}^{A}\nabla_{n}^{2}\right)|\phi\rangle,\ \ \langle\phi|V_{int}^{2}|\phi\rangle\tag{23}$$

We now use an energy-weighted sum-rule and Eqs. (11) and (12) to obtain the following equation for the monopole frequency $b_m$ (as in [9]):

$$\langle\phi|\sum_{n=1}^{A}r_{n}^{2}|\phi\rangle=\langle F_{o}|R|F_{o}\rangle=\frac{a_{o}}{b_{m}}\tag{24}$$

where $F_o\equiv F(n=0)$ and $a_o\equiv a(n=0)$.

We use the values of $\chi$, $\chi_{ms}$, $C_{mi}$, and $\Delta\varepsilon$ determined from the solution of Eq. (22), and combine Eqs. (7), (10), (13), (14), (16), (17), (19), (23) and (24) to obtain explicit expressions for the parameters $a$, $\beta_s$, $b_m$, $b_s$, $\varepsilon$, and $\chi$ in terms of $A$, $\Sigma$, monopole quantum number *n* in Eq. (17), $\chi_{mso}$, and $\chi_{mo}$.

The monopole excitation energy for the n^th excited $0^+$ state is naturally defined as follows:

$$\Delta E_{m}\equiv\frac{\hbar^{2}}{2M}\left[\varepsilon(n)-\varepsilon(n=0)\right]=\frac{\hbar^{2}b}{2M}\cdot\frac{\varepsilon(n)-\varepsilon(n=0)}{b}=\frac{\hbar\omega}{2}\cdot\frac{\varepsilon(n)-\varepsilon(n=0)}{b}\tag{25}$$

Note that $\dfrac{\varepsilon(n)-\varepsilon(n=0)}{b}$ is independent of *b* and hence $\Delta E_m$ is proportional to the oscillator frequency $\omega$ in Eq. (7).

## 4. Predictions for ${}^{4}_{2}He$, ${}^{8}_{4}Be$, ${}^{12}_{6}C$, ${}^{16}_{8}O$, ${}^{20}_{10}Ne$, ${}^{24}_{12}Mg$, and ${}^{28}_{14}Si$

In this Section, we apply the model derived in Section 3 to each of the nuclei for ${}^{4}_{2}He$, ${}^{8}_{4}Be$, ${}^{12}_{6}C$, ${}^{16}_{8}O$, ${}^{20}_{10}Ne$, ${}^{24}_{12}Mg$, and ${}^{28}_{14}Si$ and calculate the values of the model parameters $a$, $\beta_s$, $b_m$, $b_s$, $\varepsilon$, $\chi$, and $\chi_{mm}$, and the excitation energy $\Delta E_m$ in Eq. (25) of the first excited $0^+$ state. We note that we apply the model with an isotropically symmetric intrinsic state to all of these nuclei even though some of them are deformed in their ground and/or excited states because the nuclear monopole radius used is the trace of the particle quadrupole tensor distribution, which represents the overall deformation of the nucleus. The values of the parameters appearing in the model are either derived from physical arguments or determined from the model equations, and hence there are no adjustable parameters in the model.

For each of the above nuclei, we obtain the excited nuclear intrinsic state $|\phi\rangle$ by promoting a single particle from the ground state $|\phi_o\rangle$ to an orbital above the Fermi surface (as in Eq. (21)), and ensuring that the hole orbital below the Fermi surface and particle orbital above the Fermi state have the same angular momentum so that these orbitals can be coupled to zero angular momentum, and obtain zero angular momentum state $|\phi\rangle$. Furthermore, since the constraining operator $\sum_{n,j=1}^{A,3}\left(x_{nj}\right)^{-1}\partial/\partial x_{nj}$ and the residual two-body interaction $V_{\text{int}}^2$ are rotationally invariant, they do not couple states with different angular momentum. In particular, these operators have zero matrix elements between states of odd and even angular momenta. We have found it sufficiently accurate to include only $2\hbar\omega$ and $4\hbar\omega$ particle excitations in $|\phi\rangle$ since the coefficients of $4\hbar\omega$ and $6\hbar\omega$ excited states in the expansion in Eq. (21) are found to be small relative to that of $2\hbar\omega$ excited state.

Table 1 shows that the predicted excitation energy ($\Delta E_m$) agrees closely with that observed experimentally. Figures 1, 2 and 3 present the predicted variation in the model parameters, which, except for $\Delta E_m$, are rendered dimensionless. These variations are explained as follows recalling that a variation in one of the model parameters results in corresponding variations in all the other parameters because they are self-consistently determined by the model equations.

$\Delta E_m$ is a sum of n=1 and n=0 state differences in the monopole and intrinsic components of the total energy $\varepsilon$ (refer to Eq. (25). These differences are shown in Fig 1. The intrinsic energy difference is positive and increases with $A$. The monopole energy difference component is negative and decreases and more rapidly than the intrinsic energy difference as Fig 1 indicates. Therefore, $\Delta E_m$ generally decreases with $A$ as Fig 2 shows. $\Delta E_m$ also generally decreases with $A$ because it is proportional to the oscillator frequency $\omega$ (refer to Eqs. (7) and (25)) and hence to $A^{-1/3}$ as the nucleus becomes softer due to the saturation property of the short-ranged inter-nucleon force.

Starting from a certain relatively high value of $\beta_s$ (i.e., the monopole-system monopole moment defined in Eq. (15)), the expectation value $<D>\equiv\langle\phi|D|\phi\rangle$ of the intrinsic-monopole constraining dilation operator $D$ in Eq. (20) is negative and increases with $A$ and $l$ (the orbital angular momentum for a sub-shell single-particle state). Therefore, the constraining Lagrange multiplier $\chi$ must generally decrease (as Fig 2 shows) to make $< D >$ vanish as required by the constraint in Eq. (14). Fig 3 shows that $\beta_s$ generally decreases with $A$ to make $< D >$ vanish. Fig 2 shows that the mean intrinsic-system monopole dilation $a\equiv\langle\phi|\tilde{B}|\phi\rangle$, where $\tilde{B}$ is the intrinsic dilation operator defined in Eq. (5), increases with $A$ as the nuclear radius $\langle\phi|r^2|\phi\rangle$, which is proportional to $\Sigma$ in Eq. (7), increases rapidly and monotonically with $A$. Fig 3 shows that $b_m$ decreases with $A$ to comply with Eqs. (10) and (24) since $\Sigma$ increases rapidly and monotonically with $A$, until the increase in $a$ overtakes that in $\Sigma$ causing $b_m$ to start to increase. Fig 3 shows that $\chi_{mm}$ defined by Eqs. (7), (10), and (16) decreases with $A$ since $\Sigma$ increases rapidly and monotonically with $A$.

For ${}^4_2He$, the values of $\beta_s$ and $\chi$ needed to make $< D >$ vanish are high. For these values of $\beta_s$ and $\chi$, and because the last four nucleons in ${}^8_4Be$ begin to occupy the $l$=1 $1p_{3/2}$ Nilsson sub-shell orbital and hence increase the value of $< D >$, the expectation value $< D >$ is positive and relatively high in ${}^8_4Be$. Therefore, $\beta_s$ must decrease and $\chi$ must increase and more so in $n$=1 state (as Fig 2 and Fig 3 show) to make $< D >$ vanish in both $n$=0 and $n$=1 states. These changes in $\beta_s$ and $\chi$ cause the monopole excitation energy in ${}^8_4Be$ to be slightly higher than that in ${}^4_2He$ (as Fig 2 shows). For values of $\beta_s$ and $\chi$ in ${}^8_4Be$, $< D >$ in $n$=1 state is negative in ${}^{12}_6C$, and hence $\beta_s$ must increase (as Fig 3 shows) and $\chi$ must decrease (as Fig 2 shows) to reduce $< D >$ to zero in ${}^{12}_6C$. These changes in $\beta_s$ and $\chi$ result in a relatively high excitation energy for ${}^{12}_6C$ (as Fig 2 shows). For values of $\beta_s$ and $\chi$ in ${}^{12}_6C$, $< D >$ is less positive in ${}^{16}_8O$, and hence in ${}^{16}_8O$ $\beta_s$ and $\chi$ must decrease resulting in a lower excitation energy (as Fig 2 shows).

In ${}^{20}_{10}Ne$, the last four nucleons begin to occupy the $l$=2 $1d_{5/2}$ Nilsson sub-shell orbital increasing slightly the value of $< D >$. For this reason, the decrease in $\beta_s$ and $\chi$ needed to make $< D >$ vanish is less than it would have been otherwise, resulting in a slightly higher excitation energy in ${}^{20}_{10}Ne$ than that in ${}^{16}_8O$ (as Fig 2 and Fig 3 show). Thereafter, $\beta_s$ and $\chi$ and hence the excitation energy decrease monotonically with $A$ (as Fig 2 and Fig 3 show) until the nucleons in the ${}^{36}_{18}Ar$ nucleus (a case not reported in this article) begin to occupy the next Nilsson sub-shell state of non-zero angular momentum (i.e., not an $s$ orbital).

**Table 1.** Excited $0^+$ state excitation energy

| **Nucleus** | $\Delta E_m$ **(MeV)** predicted (Eq. 25) | $\Delta E_{exp}$ (MeV) **(**experiment) |
|---|---|---|
| ${}^{4}_{2}He$ | 20.147 | 20.1 |
| ${}^{8}_{4}Be$ | 20.218 | 20.2 |
| ${}^{12}_{6}C$ | 7.685 | 7.654 |
| ${}^{16}_{8}O$ | 6.054 | 6.049 |
| ${}^{20}_{10}Ne$ | 6.742 | 6.725 |
| ${}^{24}_{12}Mg$ | 6.445 | 6.432 |
| ${}^{28}_{14}Si$ | 5.011 | 4.979 |

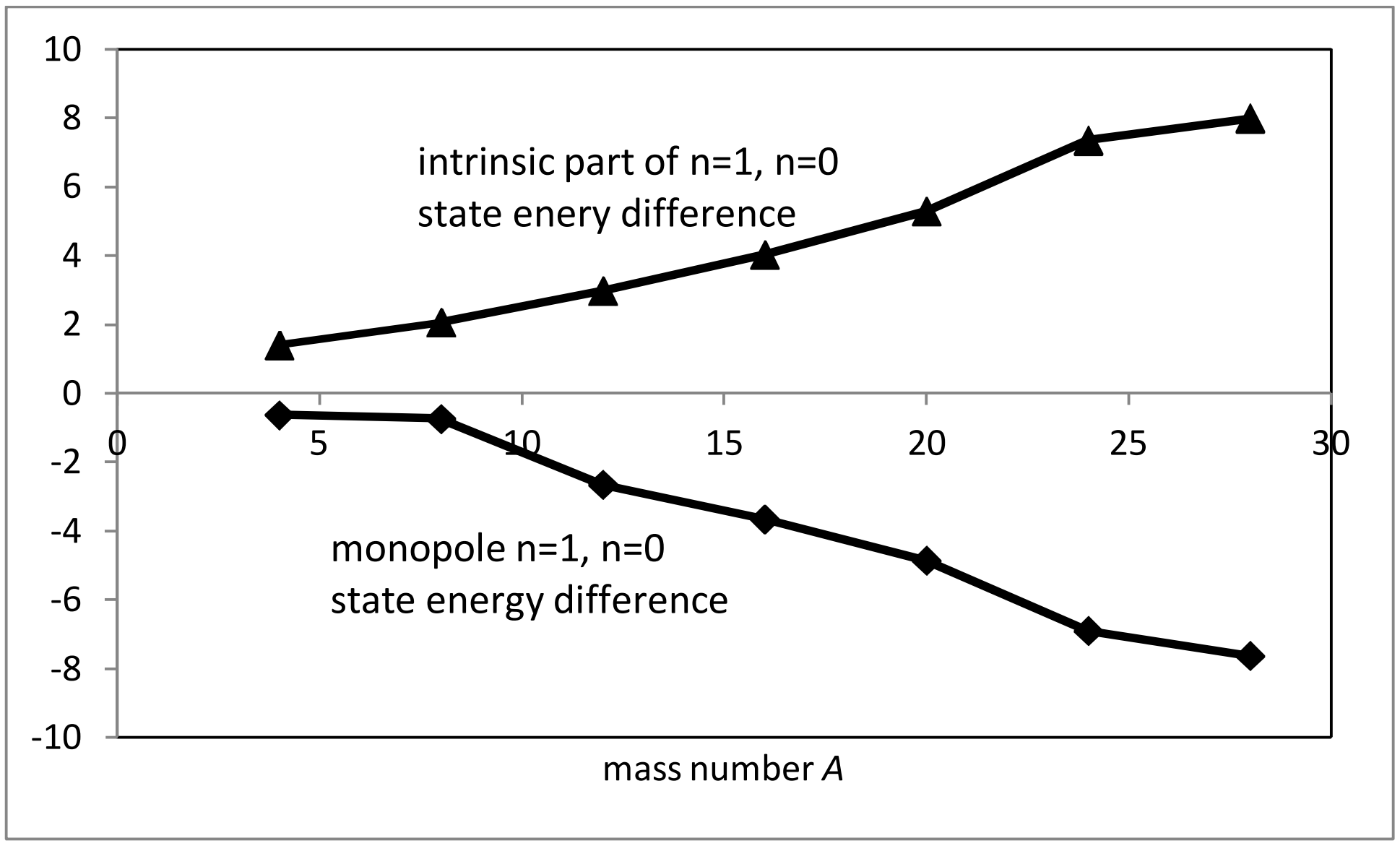

Fig 1. Variation of n=1 and n=0 state difference in monopole and intrinsic components of excitation energy ($\Delta E_m$) with mass number $A$

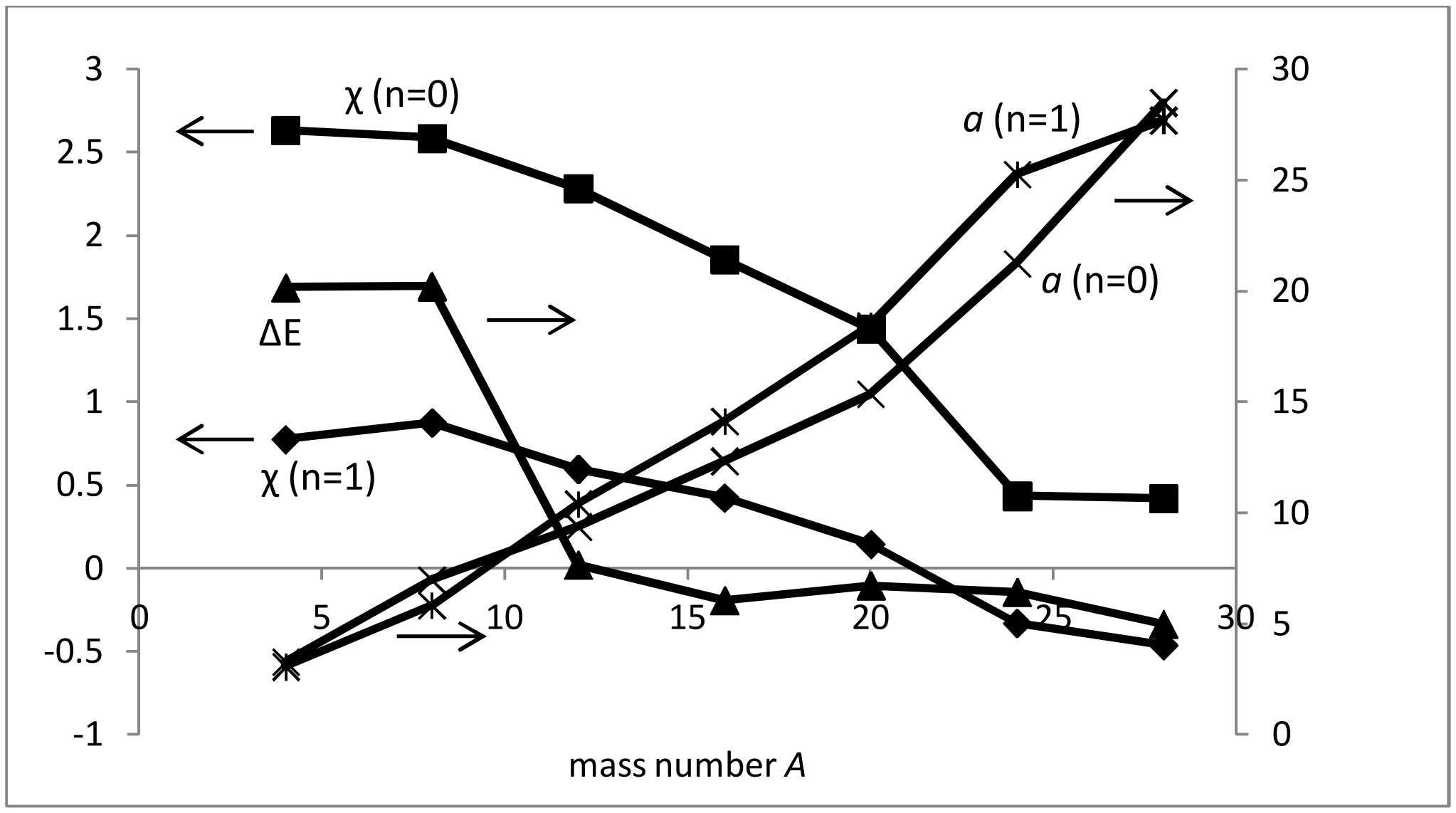

Fig 2. Variation of excitation energy ($\Delta E_m$), Lagrange multiplier ($\chi$), and mean intrinsic-system monopole dilation ($a$) with mass number $A$

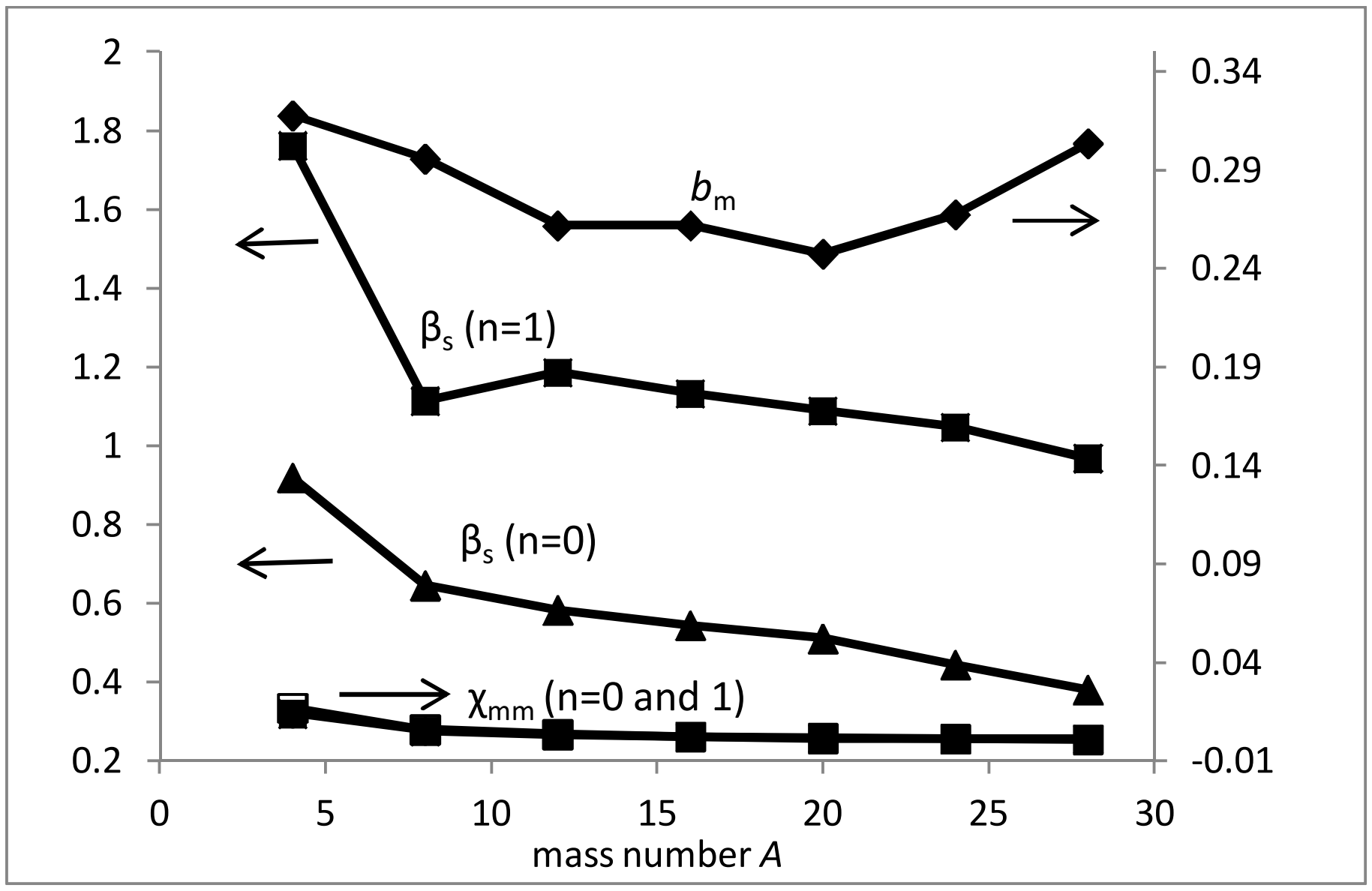

Fig 3. Variation of reduced monopole frequency ($b_m$), monopole-system monopole moment ($\beta_s$), and monopole interaction stength ($\chi_{mm}$) with mass number $A$

## 5. Concluding remarks

We have developed a quantal self-consistent cranking model for describing monopole oscillations coupled to intrinsic motion in even-even light nuclei. The derivation is based on using the nuclear radius as the collective monopole co-ordinate and a nuclear wavefunction that is a product of monopole and intrinsic wavefunctions with imposed monopole and zero-angular-momentum constraints on the intrinsic wavefunction. The intrinsic wavefunction used is isotropically symmetric to satisfy the zero angular momentum constraint even for the deformed nuclei because the non-shape related overall nuclear deformation has been extracted and represented by the monopole nuclear radius, i.e., by the trace of the trace of the particle quadrupole tensor distribution.

A transformation of the many-particle Schrodinger equation to the nuclear radius co-ordinate, partitioning of the mean-field potential and an effective two-body monopole-monopole residual interaction into their monopole and intrinsic parts, and using a variational method then results in two coupled cranking-type Schrodinger equations, one for the monopole and another for the intrinsic motions. Each of these cranking equations are time-reversal invariant since the cranking parameters, are determined self-consistently by the solutions of the equations themselves, unlike the conventional phenomenological Inglis' cranking models. (It is now apparent that the monopole and intrinsic wavefunctions in the product wavefunction are strongly coupled to each other by the two intrinsic-system constraints and by the two governing cranking equations and the cranking-equation parameters, which are determined self-consistently by the two equations themselves.)

The monopole frequency is determined using the cranking equations and an energy-weighted sum rule. The strength of the residual interaction strength is determined from a Hartree-Fock self-consistency argument. Other parameters appearing in the model are either derived from physical arguments or determined from the model equations, and hence there are no adjustable parameters in the model.

The solutions of the cranked equations and the equations associated with the cranking parameters including the equations for the constraints on the intrinsic state are determined self-consistently using Tamm-Dancoff approximation for the intrinsic state. The monopole excitation energy is then obtained from the calculated energy eigenvalue of the cranked equations.

The predicted the excitation energies for the first $0^+$ state in ${}^4_2He$, ${}^8_4Be$, ${}^{12}_6C$, ${}^{16}_8O$, ${}^{20}_{10}Ne$, ${}^{24}_{12}Mg$, and ${}^{28}_{14}Si$ are found to agree closely with the corresponding observed excitation energies. The monopole constraint imposed on the intrinsic system to suppress spurious monopole excitation in the intrinsic system is found to affect strongly the value of the monopole excitation energy. In particular, the intrinsic-system monopole moment is found to increase whenever the nucleons in a nucleus begin to occupy a new sub-shell orbital of a non-zero angular momentum (i.e., not an $s$ orbital). The resulting changes in the constraining (Lagrange) parameter and in the monopole-system monopole moment needed to suppress this spurious monopole excitation causes the monopole excitation energy to increase.